\documentclass[11pt]{article}
\usepackage{authblk}
\usepackage[colorlinks,linkcolor=blue,anchorcolor=red,citecolor=red,pdfstartview=FitH]{hyperref}
\usepackage{graphics,graphicx}
\usepackage{mathtools}
\usepackage{amsmath,amssymb,amsthm,accents,amsfonts}
\usepackage{color}
\usepackage{cite}
\usepackage{fullpage}
\usepackage{extarrows}

\numberwithin{equation}{section}
\allowdisplaybreaks

\DeclareMathAlphabet{\mathpzc}{OT1}{pzc}{m}{it}

\begin{document}

\title{\bf {Binary Darboux Transformation for the Sasa-Satsuma Equation}}
\author[1]{\bf  Jonathan J.C. Nimmo}
\author[1,2]{\bf Halis Yilmaz}
\affil[1]{School of Mathematics and Statistics, University of Glasgow, Glasgow G12 8QW, UK}
\affil[2]{Department of Mathematics, University of Dicle, 21280 Diyarbakir, Turkey}
\date{\today}

\maketitle

\begin{abstract}
The Sasa-Satsuma equation is an integrable higher-order nonlinear Schr\"odinger equation. Higher-order and multicomponent generalisations of the nonlinear Schr\"odinger equation are important in various applications, e.g., in optics. One of these equations is the Sasa-Satsuma equation. We present the binary Darboux transformations for the Sasa-Satsuma equation and then construct its quasigrammians  solutions by iterating its binary Darboux transformations. Periodic, one-soliton, two-solitons and breather solutions are given as explicit examples.
\end{abstract}

\quad{\text{\it{Keywords:}} Sasa-Satsuma equation; Binary Darboux transformation;  Quasideterminants.}

\quad{\text{\it{2010 Mathematics Subject Classification:}} 35C08, 35Q55, 37K10, 37K35}

\section{Introduction}
The celebrated nonlinear Schr\"odinger (NLS) equation
\begin{eqnarray}\label{NLS}
  iq_{t}+\frac{1}{2}q_{xx}+|q|^{2}q=0
\end{eqnarray}
 is considered to be one of the fundamental integrable equations admitting an $n-$soliton solution. It is proved integrable via the inverse scattering transform \cite{ZkSh}. The NLS equation has applications in a wide variety of physical systems such as water waves \cite{BNew, Ben, Zkh1}, plasma physics \cite{Zkh2} and nonlinear optics \cite{Has1, Has2}. This equation can be used to model short soliton pulses in optical fibres \cite{Kiv}. The basic phenomena are described by the nonlinear Schr\"odinger equation \eqref{NLS}, but as the pulses get shorter various additional effects become important. In light of this fact, Kodama and Hasegawa \cite{Kod, KodHas} developed a suitable higher-order NLS equation to take these additional effects. Their equation takes the form 
\begin{eqnarray}\label{KHeqn}
 iq_{T}+\alpha_{1}q_{XX}+\alpha_{2}|q|^{2}q+i\delta\left[\beta_{1}q_{XXX}+\beta_{2}|q|^{2}q_{X}+\beta_{3}q \left(|q|^{2}\right)_{X}\right]=0,
\end{eqnarray} 
where the $\alpha_{i}$, $\beta_{i}$ are real constants, $\delta$ is a real spectral parameter and $q$ is a complex valued function. 
The first three terms (Setting $\delta=0$ and  $\alpha_{2}=2\alpha_{1}=1$ ) form the standard nonlinear Schr\"odinger equation \eqref{NLS}. In general, the Kodama-Hasegawa higher-order NLS equation \eqref{KHeqn} is not integrable unless some restrictions are imposed on the real constants $\beta_{i}$ $(i=1,2,3)$. With appropriate choices of these real constants, the inverse scattering transform can be applied to verify integrability of the resulting equation. It is known until now, along with the NLS equation \eqref{NLS} itself, there are four cases in which integrability can be proven via inverse scattering transform. These are the Chen-Lee-Liu \cite{CLL} derivative NLS equation $(\beta_{1}:\beta_{2}:\beta_{3}=0:1:0)$, the Kaup-Newell \cite{KN} derivative NLS equation $(\beta_{1}:\beta_{2}:\beta_{3}=0:1:1)$, the Hirota \cite{Hirota} NLS equation $(\beta_{1}:\beta_{2}:\beta_{3}=1:6:0)$ and the Sasa-Satsuma \cite{SS} NLS equation $(\beta_{1}:\beta_{2}:\beta_{3}=1:6:3)$. \par 
Sasa and Satsuma \cite{SS} consider the case where $\alpha_{1}=\frac{1}{2}$ and $\alpha_{2}=1$, that is
\begin{eqnarray}\label{SSUeqn}
 iq_{T}+\frac{1}{2}q_{XX}+|q|^{2}q+i\delta\left[q_{XXX}+6|q|^{2}q_{X}+3q \left(|q|^{2}\right)_{X}\right]=0.
\end{eqnarray} 
Sasa and Satsuma \cite{SS} introduce variable transformations 
\begin{eqnarray}
 u(x,t)=q(X,T)\exp\left\{\frac{-i}{6\delta}\left(X-\frac{T}{18\delta}\right)\right\},~~t=T,~~x=X-\frac{T}{12\delta}.
\end{eqnarray}
Then, setting $\delta=1$, the equation \eqref{SSUeqn} is reduced to a complex modified KdV-type equation
\begin{eqnarray}\label{SSeqn}
 u_{t}+u_{xxx}+6|u|^{2}u_{x}+3u\left(|u|^{2}\right)_{x}=0
\end{eqnarray}
which is an equivalent version of \eqref{SSUeqn}. The equation \eqref{SSeqn} is commonly known as the Sasa-Satsuma (SS) equation, and we will denote it as such from now on. The SS equation \eqref{SSeqn} is one of the integrable extensions of the NLS equation \eqref{NLS}.  The integrability of this equation has been widely studied with various methods such as the inverse scattering scheme \cite{SS}, the Hirota's bilinear approach \cite{GNO,GKundu} and the \textit{Darboux-like} transformations \cite{Xu}.

In the present paper, we present a systematic approach to the construction of  \eqref{SSeqn} by means of  a standard binary Darboux transformation (BDT) and written in terms of quasideterminants \cite{Gelfand91,Gelfand05}. Quasideterminants have various nice properties which play important roles in constructing exact solutions of integrable systems \cite{HH,Ham,LiJon, NH, HY}. 

In this paper, we establish for the first time a standard BDT for the SS equation \eqref{SSeqn}. Our solutions for the SS equation are written in terms of quasigrammians rather than determinants. It should be emphasised that these quasigrammian solutions arise naturally from the binary Darboux transformation we present here. Furthermore, we present periodic, one-soliton, two-solitons and breather solutions for the SS equation.

This paper is organized as follows. In Section \ref{Qdet} below, we give a brief review on quasideterminants. In Section \ref{EVss}, we construct a $3\times 2$ eigenfunction and corresponding constant $2\times 2$ square matrix for the eigenvalue problems of the SS equation \eqref{SSeqn} via a symmetry matrix.
In Section \ref{bDTss}, we state a standard binary Darboux theorem for the Sasa-Satsuma system. 
In Section \ref{bDTrss}, we review the reduced binary Darboux transformations for the SS equation, which can be considered as a dimensional reduction from $(2+1)$ to $(1+1)$ dimensions.
In Sections \ref{QGss}, we present the quasigrammian solutions of the SS equation by using the binary Darboux transformation. Here, the quasigrammians are written in terms of solutions of linear eigenvalue problems. In Section \ref{PSss}, periodic, one-soliton, two-solitons and breather solutions of the SS equation are given for both zero and non-zero seed solutions. The conclusion is given in the final Section \ref{ConSS}.

\subsection{Quasideterminants}\label{Qdet}

In this short section we recall some of the key elementary properties of quasideterminants. The reader is referred to the original papers \cite{Gelfand91,Gelfand05} for a more detailed and general treatment.

Quasideterminants were introduced by Gelfand and Retakh in \cite{Gelfand91} as a natural generalisation of the determinant to matrices with noncommutative entries. Many equivalent definitions of quasideterminants exist, one such being a recursive definition involving inverse minors. Let $M=(m_{ij})$ be an $n\times n$ matrix  with entries over an, in general non-commutative, ring $\mathpzc{R}$. Then the quasideterminants of $M$ for $i,j=1,\dots,n$ are defined by
\begin{equation}\label{expan}
 |M|_{ij}=m_{ij}-r_i^j\left(M^{ij}\right)^{-1}c_j^i ,
\end{equation}
where $r_i^j$ is the row vector obtained from $i^{th}$ row of $M$ with the $j^{th}$ element removed, $c_j^i$ is the column vector obtained from $j^{th}$ column of $M$ with the $i^{th}$ element removed and $M^{ij}$ is the $(n-1)\times (n-1)$ submatrix obtained by deleting the $i^{th}$ row and the $j^{th}$ column from $M$. Quasideterminants can also be denoted by boxing the entry about which the expansion is made
\begin{align}
 |M|_{ij}=
         {\left|\begin{array}{cc}
          M^{ij}& c_j^i\\ r_i^j & \boxed{m_{ij}}
         \end{array}\right|}.
\end{align}
If the entries in $M$ happen to commute, then the quasideterminant $|M|_{ij}$ can be expressed as a ratio of determinants
\begin{equation}\label{comdet}
 |M|_{ij}=(-1)^{i+j}\frac{\det M}{\det M^{ij}}.
\end{equation}

In this paper, we will consider only quasideterminants that are expanded about a term in the last column, most usually the last entry. For example considering a block matrix $M=\left(\begin{array}{cc} A& B\\ C & d \end{array}\right)$, where $A$ is an invertible (square) matrix over $\mathpzc{R}$ of arbitrary size and $B$, $C$ are column and row vectors 
over $\mathpzc{R}$ of compatible lengths, respectively, and $d\in\mathpzc{R}$, the quasideterminant of $M$ is expanded about $d$ is 
\begin{align}
         \left|\begin{array}{cc}
          A& B\\ C & \boxed{d}
         \end{array}\right|
    =d-CA^{-1}B.
\end{align}
Later we will use the following invariance of quasideterminants which follows immediately from their definition. Let $\alpha$ and $\beta$ be invertible matrices of the same dimensions as $A$. Then
\begin{align}\label{qinv}
         \left|\begin{array}{cc}
          \alpha A\beta& \alpha B\\ C\beta & \boxed{d}
         \end{array}\right|=
         \left|\begin{array}{cc}
          A& B\\ C & \boxed{d}
         \end{array}\right|.
\end{align}

\section{The eigenvalue problems for the Sasa-Satsuma Equation}\label{EVss}

The Lax pair \cite{SS} for the Sasa-Satsuma equation \eqref{SSeqn} is given by 
\begin{eqnarray}
 L&=&\partial_x+J\lambda+R\label{LaxL}\\
 M&=&\partial_t+4J\lambda^3+4R\lambda^2-2Q\lambda+W,\label{LaxM}
\end{eqnarray}
where $J$, $R$, $Q$ and $W$ are $3\times 3$ matrices such that
\begin{eqnarray}\label{JRQ}
  J={\left(\begin{array}{ccc}
          i & 0 & 0 \\ 0 & i & 0 \\ 0 & 0 & -i
  \end{array}\right)},\hspace{0.3cm}
   R={\left(\begin{array}{ccc}
          0& 0 & -u \\ 0 & 0 & -u^{*}\\ u^{*} & u & 0
  \end{array}\right)},\hspace{0.3cm}
   Q={\left(\begin{array}{ccc}
          |u|^{2}&u^{2}&u_{x}\\{u^{*}}^{2}&|u|^{2}&u^{*}_{x}\\u^{*}_{x}&u_{x}&-2|u|^{2}
  \end{array}\right)i}
\end{eqnarray}
and
\begin{eqnarray}\label{W}
 W=\left(\begin{array}{ccc}
         u^{*}u_{x}-uu^{*}_{x} & 0 & 4u|u|^{2}+u_{xx}\\
           0 & uu^{*}_{x}-u^{*}u_{x} &  4u^{*}|u|^{2}+u^{*}_{xx} 
           \\ -4u^{*}|u|^{2}-u^{*}_{xx}&-4u|u|^{2}-u_{xx}& 0
  \end{array}\right).
\end{eqnarray}
Here $\lambda$ is a spectral parameter and asterisk denotes complex conjugate. 
It can be seen that the potential matrix $R$ in \eqref{JRQ} has two symmetry properties \cite{KY, YK}. 
One is that it is skew-Hermitian: $R^{\dagger}=-R$.  The other one is that $S R S^{-1}=R^{*}$, where 
\begin{eqnarray}\label{symS}
 S=S^{-1}=\left(\begin{array}{ccc}
        0 & 1 & 0\\ 1 & 0 &  0\\ 0 & 0 & 1
  \end{array}\right).
\end{eqnarray}
Let $\phi=(\phi_{1}, \phi_{2}, \phi_{3})^{T}$ be a vector eigenfunction for \eqref{LaxL}-\eqref{LaxM} for eigenvalue $\lambda$. Using the second symmetry, it is easy to see that $S\phi^{*}=(\phi_{2}^{*}, \phi_{1}^{*}, \phi_{3}^{*})^{T}$ 
is another eigenfunction, for eigenvalue $-\lambda^{*}$. Using these vector eigenfunctions we can define a $3\times 2$ matrix eigenfunction $\theta$ with $2\times 2$ eigenvalue $\Lambda$
\begin{eqnarray}\label{thetaLambda}
 \theta=\left(\begin{array}{ccc}
        \phi_{1} & \phi_{2}^{*}\\ \phi_{2} & \phi_{1}^{*}\\ \phi_{3} & \phi_{3}^{*}
  \end{array}\right), \hspace{0.3cm}
  \Lambda=\left(\begin{array}{cc}\lambda & 0\\ 0 & -\lambda^{*}\end{array}\right),
\end{eqnarray}
satisfying
\begin{eqnarray}
  &&\theta_{x}+J\theta \Lambda+R\theta=0,\label{LaxEqnX}\\
 && \theta_{t}+4J\theta \Lambda^{3}+4R\theta \Lambda^{2}-2Q\theta\Lambda+W\theta=0.\label{LaxEqnT}
\end{eqnarray}

\section{Darboux transformations and Dimensional reductions}

\subsection{Darboux transformation}
Let us consider the linear operators
\begin{equation}\label{LM}
 L=\partial_x+\sum_{i=0}^nu_i\partial_y^i,\hspace{0.5cm}M=\partial_t+\sum_{i=0}^nv_i\partial_y^i,
\end{equation}
where $u_{i}, v_{i} \in \mathpzc{R}$, where $\mathpzc{R}$ is a ring, in general non-commutative.
The standard approach to Darboux transformations \cite{Darboux, Matveev, MS} involves a gauge operator $G_{\theta}=\theta\partial_{y}\theta^{-1}$, where 
$\theta=\theta(x,y,t)$ is a solution to a linear system 
\begin{eqnarray}
 L(\phi)=M(\phi)=0,
\end{eqnarray}
where $\phi$ is any eigenfunction of $L$ and $M$, has the property of leaving the above linear problems invariant:
\begin{eqnarray}
 \tilde{L}(\tilde{\phi})=\tilde{M}(\tilde{\phi})=0,
\end{eqnarray}
where $\tilde{\phi}=G_{\theta}(\phi)$ is an eigenfunction of $\tilde{L}$ and $\tilde{M}$ \eqref{LM} and the linear operators
$\tilde{L}=G_{\theta}LG_{\theta}^{-1}$ and $ \tilde{M}=G_{\theta}MG_{\theta}^{-1}$ have the same forms as $L$ and $M$:
\begin{equation}
 \tilde{L}=\partial_x+\sum_{i=0}^n\tilde{u}_i\partial_y^i,\hspace{0.5cm}\tilde{M}=\partial_t+\sum_{i=0}^n\tilde{v}_i\partial_y^i.
\end{equation}

\subsection{Binary Darboux transformation}\label{bDTss}

Let $G_{\theta}$ and $G_{\hat{\theta}}$ be two standard Darboux transformations map two linear operators $L$ and $\hat{L}$ onto a common linear operator $\tilde{L}$ such that
\begin{eqnarray}
 \tilde{L}=G_{\theta}LG_{\theta}^{-1}=G_{\hat{\theta}}\hat{L}G_{\hat{\theta}}^{-1}.
\end{eqnarray}
Then one may define a binary Darboux transformation \cite{MS, NGO} $B_{\theta,\hat{\theta}}=G_{\hat{\theta}}^{-1}G_{\theta}$ such that 
$\hat{L}=B_{\theta,\hat{\theta}}L B_{\theta,\hat{\theta}}^{-1}$.  In order to define $G_{\hat{\theta}}$ one needs an eigenfunction of $\hat{L}$. 
This problem can be got around by using the formal adjoint operator $L^{\dagger}$ constructed according to the rule $\left(a\partial_{y}^{i}\right)^{\dagger}=\left(-\partial_{y}\right)^{i}a^{\dagger}$, where $a^{\dagger}$ denotes the Hermitian conjugate of $a$.  If $L(\phi)=L(\theta)=0$ and $L^{\dagger}(\psi)=L^{\dagger}(\rho)=0$, we derive the eigenfunction $\hat{\theta}$ as
\begin{eqnarray}
 \hat{\theta}=\theta\Omega(\theta,\rho)^{-1},
\end{eqnarray}
where the eigenfunction potential $\Omega$ is defined such that $\Omega(\theta,\rho)_{y}=\rho^{\dagger}\theta$.
We can then construct the binary Darboux transformation explicitly as
\begin{eqnarray}
 B_{\theta,\rho}=I-\theta\Omega(\theta,\rho)^{-1}\partial_{y}^{-1}\rho^{\dagger}
\end{eqnarray}
with  adjoint 
\begin{eqnarray}
 B_{\theta,\rho}^{-\dagger}=I-\rho\Omega(\theta,\rho)^{-\dagger}\partial_{y}^{-1}\theta^{\dagger}.
\end{eqnarray}

It should be pointed out that  the above formulas obtained  for $B_{\theta,\rho}$ and $B_{\theta,\rho}^{-\dagger}$, the transformation makes sense for any $m \times n$ matrices $\theta$ and $\rho$ such that $L(\theta)=L^{\dagger}(\rho)=0$, not just square $m\times m$ matrices. Here, only the eigenfunction potential $\Omega(\theta,\rho)$ must be an invertible square matrix.

Suppose $\phi_{[1]}=\phi$ is a general eigenfunction of the operators $L_{[1]}=L$, $M_{[1]}=M$ and $\psi_{[1]}=\psi$ a general eigenfunction of the adjoint Lax operators $L^{\dagger}_{[1]}=L^{\dagger}$, $M^{\dagger}_{[1]}=M^{\dagger}$, where $L$ and $M$ are given by \eqref{LM}. We then define  the binary Darboux transformations of  the eigenfunctions ${\phi}$ and $\psi$ as
\begin{eqnarray}
\phi_{[2]}&=&B_{\theta,\rho}\left(\phi_{[1]}\right)=\phi_{[1]}-\theta_{[1]}\Omega\left(\theta_{[1]},\rho_{[1]}\right)^{-1}\Omega\left(\phi_{[1]},\rho_{[1]}\right),\\
\psi_{[2]}&=&B_{\theta,\rho}^{-\dagger}\left(\psi_{[1]}\right)=\psi_{[1]}-\rho_{[1]}\Omega\left(\theta_{[1]},\rho_{[1]}\right)^{-\dagger}\Omega\left(\theta_{[1]},\psi_{[1]}\right)^{\dagger},
\end{eqnarray}
with
\begin{eqnarray}
 \theta_{[2]}=\phi_{[2]}\vert_{\phi\rightarrow \theta_{2}},\hspace{0.5cm}\rho_{[2]}=\psi_{[2]}\vert_{\psi\rightarrow \rho_{2}}.
\end{eqnarray}
After $n\geqslant 1$ iterations, the $n$th binary Darboux transformation is given by
\begin{eqnarray}
\phi_{[n+1]}&=&B_{\theta,\rho}\left(\phi_{[n]}\right)=\phi_{[n]}-\theta_{[n]}\Omega\left(\theta_{[n]},\rho_{[n]}\right)^{-1}\Omega\left(\phi_{[n]},\rho_{[n]}\right),\\
\psi_{[n+1]}&=&B_{\theta,\rho}^{-\dagger}\left(\psi_{[n]}\right)=\psi_{[n]}-\rho_{[n]}\Omega\left(\theta_{[n]},\rho_{[n]}\right)^{-\dagger}\Omega\left(\theta_{[n]},\psi_{[n]}\right)^{\dagger},
\end{eqnarray}
with
\begin{eqnarray}
 \theta_{[n]}=\phi_{[n]}\vert_{\phi\rightarrow \theta_{n}},\hspace{0.5cm}\rho_{[n]}=\psi_{[n]}\vert_{\psi\rightarrow \rho_{n}},
\end{eqnarray}
where  $L(\theta_{i})=M(\theta_{i})=0$ and  $L^{\dagger}(\rho_{i})=M^{\dagger}(\rho_{i})=0$. Using the notation $\Theta=\left(\theta_{1},\ldots,\theta_{n}\right)$ and $P=\left(\rho_{1},\ldots,\rho_{n}\right)$, we have
\begin{eqnarray}
\phi_{[n+1]}=
\begin{vmatrix}\Omega(\Theta, P) & \Omega(\phi, P)\\\Theta & \boxed{\phi}\end{vmatrix},\hspace{0.2cm}
\psi_{[n+1]}=\begin{vmatrix}\Omega(\Theta, P)^{\dagger} & \Omega(\Theta,\psi)^{\dagger}\\P & \boxed{\psi}\end{vmatrix},
\end{eqnarray}
and
\begin{eqnarray}
\Omega(\phi_{[n+1]},\psi_{[n+1]})=
\begin{vmatrix}\Omega(\Theta, P) & \Omega(\phi, P)\\ \Omega(\Theta, \psi) & \boxed{\Omega(\phi, \psi)}\end{vmatrix}.
\end{eqnarray}

\subsection{Dimensional reductions of the binary Darboux transformation}\label{bDTrss}

Here, we describe a reduction of the binary Darboux transformation from $(2+1)$ to $(1+1)$ dimensions. We choose to  eliminate the $y$-dependence by employing a `separation of variables' technique. The reader is referred to the paper \cite{NGO} for a more detailed treatment. We make the ans\"atze
\begin{eqnarray}
 \phi &=&\phi^r(x,t)e^{\lambda y},\hspace{0.5cm}\theta =\theta^r(x,t)e^{\Lambda y},\\
 \psi &=&\psi^r(x,t)e^{\mu y},\hspace{0.5cm}\rho =\rho^r(x,t)e^{\Pi y},
\end{eqnarray}
where $\lambda, \mu$ are constant scalars and $\Lambda, \Pi$ are $N \times N$ constant matrices and
the superscript $r$ denotes reduced functions, independent of $y$.
Hence in the dimensional reduction we obtain $\partial_y^{i}\left(\phi\right)=\lambda^i\phi$ and
$\partial_y^{i}\left(\theta \right)=\theta \Lambda^i$ and so the operator $L$ in \eqref{LM} becomes
 \begin{eqnarray}
  L^r&=&\partial_x+\sum_{i=0}^n u_i\lambda^i,
 \end{eqnarray}
where $\theta^r$ is a matrix eigenfunction of $L^r$ such that $L^r\left(\theta^r\right)=0$, with $\lambda$ replaced by the matrix $\Lambda$, that is,
\begin{equation}
	\theta^r_x+\sum_{i=0}^n u_i\theta^r\Lambda^i=0.
\end{equation}
It follows that the $y-$dependence of the potential $\Omega$ can also be made explicit by setting $\Omega(\theta,\rho)=e^{\Pi^{\dagger}y}\Omega^{r}\left(\theta^{r},\rho^{r}\right)e^{\Lambda y}$ and 
$\Omega(\phi,\rho)=e^{\left(\Pi^{\dagger}+\lambda I\right)y}\Omega^{r}\left(\phi^{r},\rho^{r}\right)$. Then the dimensionally reduced binary Darboux transformations are written as 
 \begin{eqnarray}
 B_{\theta^{r},\rho^{r}}&=&I-\theta^{r}\Omega^{r}\left(\theta^{r},\rho^{r}\right)^{-1}\left(\Pi^{\dagger}+\lambda I\right)^{-1}\rho^{r\dagger},\\
  B_{\theta^{r},\rho^{r}}^{-\dagger}&=&I-\rho^{r}\Omega^{r}\left(\theta^{r},\rho^{r}\right)^{-\dagger}\left(\Lambda^{\dagger}+\mu I\right)^{-1}\theta^{r\dagger},
 \end{eqnarray}
where $\Omega^{r}$ is an algebraic potential satisfying the following conditions 
\begin{eqnarray}
 \Pi^{\dagger}\Omega^{r}\left(\theta^{r},\rho^{r}\right)+\Omega^{r}\left(\theta^{r},\rho^{r}\right)\Lambda&=&\rho^{r\dagger}\theta^{r},\label{relation1}\\
 \left(\Pi^{\dagger}+\lambda I\right)\Omega^{r}\left(\phi^{r},\rho^{r}\right)&=&\rho^{r\dagger}\phi^{r},\\
\Omega^{r}\left(\theta^{r},\psi^{r}\right)\left(\Lambda+\mu^{\dagger}I\right)&=&\psi^{r\dagger}\theta^{r}.
\end{eqnarray} 
The transformed operators 
\begin{eqnarray}
L_{[n+1]}^{r}=B_{\theta^{r},\rho^{r}}L_{[n]}^{r}B_{\theta^{r},\rho^{r}}^{-1},\\
M_{[n+1]}^{r}=B_{\theta^{r},\rho^{r}}M_{[n]}^{r}B_{\theta^{r},\rho^{r}}^{-1}
\end{eqnarray}
have generic eigenfunctions and adjoint eigenfunctions
\begin{eqnarray}\label{SSphi}
\phi_{[n+1]}^{r}&=&\phi_{[n]}^{r}-\theta_{[n]}^{r}\Omega^{r}\left(\theta_{[n]}^{r},\rho_{[n]}^{r}\right)^{-1}\Omega^{r}\left(\phi_{[n]}^{r},\rho_{[n]}^{r}\right),\\
\psi_{[n+1]}^{r}&=&\psi_{[n]}^{r}-\rho_{[n]}^{r}\Omega^{r}\left(\theta_{[n]}^{r},\rho_{[n]}^{r}\right)^{-\dagger}\Omega^{r}\left(\theta_{[n]}^{r},\psi_{[n]}^{r}\right)^{\dagger},
\end{eqnarray}
with
\begin{eqnarray}
 \theta^{r}_{[n]}=\phi^{r}_{[n]}\vert_{\phi^{r}\rightarrow \theta_{n}},\hspace{0.5cm}\rho^{r}_{[n]}=\psi^{r}_{[n]}\vert_{\psi^{r}\rightarrow \rho^{r}_{n}}.
\end{eqnarray}
From now on, for notational simplicity, we omit the superscript $r$.

\section{Quasi-Grammian solutions of the Sasa-Satsuma equation}\label{QGss}

In this section we determine the effect of the BDT 
$B_{\theta,\rho}=I-\theta\Omega(\theta,\rho)^{-1}\left(\Pi^{\dagger}+\lambda I\right)^{-1}\rho^{\dagger}$
on the operator $L=\partial_x+J\lambda+R$ given by \eqref{LaxL} with $\theta$  an eigenfunction of $L$ and $\rho$ an eigenfunction of $L^{\dagger}$. Corresponding results hold for the operator $M$ given by \eqref{LaxM} and its corresponding adjoint $M^{\dagger}$. 
In the operator $L$, the matrix coefficients $J$ and $R$ are both skew-Hermitian. This leads us to observe the relation $L+L^{\dagger}=0$ in which we can let $\rho=\theta$ and $\psi=\phi$.

The operator $L$ is transformed to a new operator $\hat{L}$ such that
\begin{eqnarray}
 \hat{L}=B_{\theta}L B_{\theta}^{-1},
\end{eqnarray}
where  $B_\theta=I-\theta\Omega(\theta,\theta)^{-1}\left(\Pi^{\dagger}+\lambda I\right)^{-1}\theta^{\dagger}$.

We find that 
\begin{eqnarray}\label{bdtR}
 \hat{R}=R+\left[J,\theta \Omega(\theta,\theta)^{-1}\theta^{\dagger}\right].
\end{eqnarray}
For notational convenience, we introduce a $3 \times 3$ matrix $P=\left(p_{ij}\right)$ such that $R=[J,P]$, and hence
\begin{eqnarray}\label{solP}
  P=\frac{1}{2i}\begin{pmatrix}
    p_{11} & p_{12} & u \\ 
    p_{21} &  p_{22} & u^* \\ 
    u^* & u & p_{33}
  \end{pmatrix}.
\end{eqnarray}
From \eqref{bdtR}, since $R=[J,P]$, it follows that
\begin{eqnarray}\label{SSP}
 \hat{P}=P-\theta \Omega(\theta,\theta)^{-1}\theta^{\dagger},
\end{eqnarray}
in which the relation \eqref{relation1} is now
\begin{eqnarray}
\Omega(\theta,\theta) \Lambda-\Lambda^{\dagger}\Omega(\theta,\theta) =\theta^{\dagger}\theta
\end{eqnarray}
with $\Pi=-\Lambda$, where the eigenfunction $\theta$ and the diagonal constant matrix $\Lambda$ are given by \eqref{thetaLambda}.

Let $P_{[1]}=P$,  $P_{[2]}=\hat{P}$, $\theta_{[1]}=\theta_{1}=\theta$ and $\Lambda_{1}=\Lambda$ so that $\Lambda_{1}=\text{diag}\left(\lambda_{1},-\lambda_{1}^{*}\right)$. Then, the solution \eqref{SSP} can be rewritten  as
\begin{eqnarray}
  P_{[2]}=P_{[1]}-\theta_{[1]} \Omega(\theta_{[1]},\theta_{[1]})^{-1}\theta_{[1]}^{\dagger}.
\end{eqnarray}
After $n$ repeated applications of the reduced binary Darboux transformation $B_{\theta}$, we have
\begin{eqnarray}
  P_{[n+1]}=P_{[n]}-\theta_{[n]} \Omega(\theta_{[n]},\theta_{[n]})^{-1}\theta_{[n]}^{\dagger}
\end{eqnarray}
with $ \theta_{[n]}=\phi_{[n]}\vert_{\phi\rightarrow \theta_{n}}$, where the general eigenfunction $\phi_{[n]}$ is given by \eqref{SSphi} as
\begin{eqnarray}
 \phi_{[n+1]}=\phi_{[n]}-\theta_{[n]}\Omega(\theta_{[n]},\theta_{[n]})^{-1}\Omega(\phi_{[n]},\theta_{[n]}).
\end{eqnarray}
Let $\phi_{1},\ldots,\phi_{n}$ be a particular set of eigenfunctions of  the linear operators $L$, $M$ given by \eqref{LaxL}$-$\eqref{LaxM}, and define
$\Theta=\left(\theta_{1},\ldots,\theta_{n}\right)$ for the $3\times 2$ matrices $\theta_{i}$ $\left(i=1,\ldots,n\right)$ such that
\begin{eqnarray}
 \theta_{i}=\left(\begin{array}{ccc}
        \phi_{3i-2} & \phi_{3i-1}^{*}\\ \phi_{3i-1} & \phi_{3i-2}^{*}\\ \phi_{3i} & \phi_{3i}^{*}
 \end{array}\right).
\end{eqnarray}
We express $P_{[n+1]}$ and $\phi_{[n+1]}$ in quasi-Grammian forms as
\begin{eqnarray}
P_{[n+1]}= P+
\begin{vmatrix}
  \Omega(\Theta,\Theta) & \Theta^{\dagger}\\
  \Theta & \boxed{\begin{matrix}0&0 & 0\\0&0&0\\0&0&0\end{matrix}}
 \end{vmatrix},\hspace{0.2cm}
 \phi_{[n+1]}=\begin{vmatrix}\Omega(\Theta,\Theta) & \Omega(\phi,\Theta)\\
 \Theta & \boxed{\phi}\end{vmatrix}.
\end{eqnarray}
In order to express the quasi-Grammian solution $P_{[n+1]}$ explicitly in terms of the variable $u$, we define the $3 \times 2n$ matrix $\Theta$ as
\begin{eqnarray}\label{Thetapot}
 \Theta=\begin{pmatrix} \Phi_{1} \\ \Phi_{2}\\ \Phi_{3} \end{pmatrix},
\end{eqnarray}
where $\Phi_{1}$, $\Phi_{2}$ and $\Phi_{3}$ denote the row vectors $\left(\phi_{1},\phi_{2}^{*},\ldots,\phi_{3n-2},\phi_{3n-1}^{*}\right)$, $\left(\phi_{2},\phi_{1}^{*},\ldots,\phi_{3n-1},\phi_{3n-2}^{*}\right)$ and $\left(\phi_{3},\phi_{3}^{*},\ldots,\phi_{3n},\phi_{3n}^{*}\right)$ respectively. Thus, we obtain

\begin{eqnarray}\label{pn1}
 P_{[n+1]}=P+
 \begin{pmatrix}
 \begin{vmatrix}\Omega(\Theta,\Theta) & \Phi_{1}^{\dagger}\\ \Phi_{1} & \boxed{0}\end{vmatrix} &
 \begin{vmatrix}\Omega(\Theta,\Theta) & \Phi_{2}^{\dagger}\\ \Phi_{1} & \boxed{0}\end{vmatrix} &
 \begin{vmatrix}\Omega(\Theta,\Theta) & \Phi_{3}^{\dagger}\\ \Phi_{1} & \boxed{0}\end{vmatrix}\\\\
 \begin{vmatrix}\Omega(\Theta,\Theta) & \Phi_{1}^{\dagger}\\ \Phi_{2} & \boxed{0}\end{vmatrix} &
 \begin{vmatrix}\Omega(\Theta,\Theta) & \Phi_{2}^{\dagger}\\ \Phi_{2}& \boxed{0}\end{vmatrix} &
 \begin{vmatrix}\Omega(\Theta,\Theta) & \Phi_{3}^{\dagger}\\ \Phi_{2} & \boxed{0}\end{vmatrix}\\\\
 \begin{vmatrix}\Omega(\Theta,\Theta) & \Phi_{1}^{\dagger}\\ \Phi_{3} & \boxed{0}\end{vmatrix} &
 \begin{vmatrix}\Omega(\Theta,\Theta) & \Phi_{2}^{\dagger}\\ \Phi_{3} & \boxed{0}\end{vmatrix} &
 \begin{vmatrix}\Omega(\Theta,\Theta) & \Phi_{3}^{\dagger}\\ \Phi_{3} & \boxed{0}\end{vmatrix}
 \end{pmatrix}.
\end{eqnarray}
By substituting \eqref{solP} into \eqref{pn1}, we have quasi-Grammian expressions for $u$ and $u^{*}$, namely
\begin{eqnarray}
 u_{[n+1]}=u+2i\begin{vmatrix}\Omega(\Theta,\Theta) & \Phi_{3}^{\dagger}\\ \Phi_{1} & \boxed{0}\end{vmatrix},\label{SSqGr}\\
                 =u+2i\begin{vmatrix}\Omega(\Theta,\Theta) & \Phi_{2}^{\dagger}\\ \Phi_{3} & \boxed{0}\end{vmatrix},\label{SSqGr2}\\
 u_{[n+1]}^{*}=u^{*}+2i\begin{vmatrix}\Omega(\Theta,\Theta) & \Phi_{1}^{\dagger}\\ \Phi_{3} & \boxed{0}\end{vmatrix},\label{SSqGr3}\\
                 =u^{*}+2i\begin{vmatrix}\Omega(\Theta,\Theta) & \Phi_{3}^{\dagger}\\ \Phi_{2} & \boxed{0}\end{vmatrix}.\label{SSqGr4}
\end{eqnarray}
These form a quasi-Grammian solution $u$ of the Sasa-Satsuma equation \eqref{SSeqn} and its complex conjugate, However, it is necessary to show that these four expressions are consistent. That is, that the expressions on the right hand sides of \eqref{SSqGr}--\eqref{SSqGr2} and \eqref{SSqGr3}--\eqref{SSqGr4} are equal and that the pairs are indeed complex conjugates. The proof of this is given in Section~\ref{prf}. 

\subsection{Proof of consistency}\label{prf}
In the expressions \eqref{SSqGr}--\eqref{SSqGr4}, the potential $\Omega(\Theta,\Theta)$ is a $2n\times 2n$ matrix satisfying the relation
\begin{eqnarray}
 \Omega(\Theta,\Theta)\Lambda-\Lambda^{\dagger}\Omega(\Theta,\Theta)=\Theta^{\dagger}\Theta,
\end{eqnarray}
where $\Lambda$ is $2n\times 2n$ constant matrix such that $\Lambda=\text{diag}\left(\Lambda_{1},\ldots,\Lambda_{n}\right)$. Solving this relation for $\Omega(\Theta,\Theta)$, we obtain the explicit expression
\begin{eqnarray}\label{PotOmg}
\Omega\left(\Theta,\Theta\right)= 
 \begin{pmatrix}\Omega(\theta_{1},\theta_{1}) & \Omega(\theta_{2},\theta_{1})&\ldots &\Omega(\theta_{n},\theta_{1})\\
  \Omega(\theta_{1},\theta_{2}) & \Omega(\theta_{2},\theta_{2})&\ldots &\Omega(\theta_{n},\theta_{2})\\
  \vdots & \vdots & & \vdots\\
  \Omega(\theta_{1},\theta_{n}) & \Omega(\theta_{2},\theta_{n})&\ldots &\Omega(\theta_{n},\theta_{n})
  \end{pmatrix},
\end{eqnarray}
where $\Omega(\theta_{i},\theta_{j})$ is $2\times 2$ potential satisfying the relation
\begin{eqnarray}
 \Omega(\theta_{i},\theta_{j})\Lambda_{i}-\Lambda_{j}^{\dagger}\Omega(\theta_{i},\theta_{j})=\theta_{j}^{\dagger}\theta_{i},
\end{eqnarray}
where $\Lambda_{k}=\text{diag}\left(\lambda_{k},-\lambda_{k}^{*}\right)$ and $i,j,k\in \{1,2,\ldots,n\}$. It follows from this relation that the potential $\Omega$ can be written explicitly as
\begin{eqnarray}\label{Omega1}
 \Omega(\theta_{i},\theta_{j})=
 \begin{pmatrix}
    F_{ij} & -G_{ij}^{*}\\
    G_{ij} & -F_{ij}^{*}
 \end{pmatrix},
\end{eqnarray}
where  $F_{ij}=F_{ij}(x,t,\lambda_{i},\lambda_{j})$ and $G_{ij}=G_{ij}(x,t,\lambda_{i},\lambda_{j})$ are the scalar functions
\begin{eqnarray}
\label{F}
 F_{ij}&=&\frac{1}{\lambda_{i}-\lambda_{j}^{*}}\left(\phi_{3i-2}\phi_{3j-2}^{*}+\phi_{3i-1}\phi_{3j-1}^{*}+\phi_{3i}\phi_{3j}^{*}\right),\\
 \label{G}
 G_{ij}&=&\frac{1}{\lambda_{i}+\lambda_{j}}\left(\phi_{3i-2}\phi_{3j-1}+\phi_{3i-1}\phi_{3j-2}+\phi_{3i}\phi_{3j}\right).
\end{eqnarray}
Here we observe that $F_{ij}$ and $G_{ij}$ are such that $F_{ij}^{*}=-F_{ji}$ and $G_{ij}=G_{ji}$, for $i, j=1,\ldots,n$.
Then the $2\times2$ potentials $\Omega(\theta_{i},\theta_{j})$ satisfy the symmetry condition
\begin{eqnarray}
 \Omega(\theta_{i},\theta_{j})+\Omega(\theta_{j},\theta_{i})^{\dagger}=0,
\end{eqnarray}
and the $2n\times2n$ matrix potential $\Omega(\Theta,\Theta)$, as given by \eqref{PotOmg}, is skew-adjoint,
\begin{equation}\label{skew adjoint}
 \Omega(\Theta,\Theta)+\Omega(\Theta,\Theta)^{\dagger}=0.
\end{equation}
Now it is readily seen that
\begin{align}
 \begin{vmatrix}\Omega(\Theta,\Theta) & \Phi_{3}^{\dagger}\\ \Phi_{1} & \boxed{0}\end{vmatrix}^*
 &=
 \begin{vmatrix}\Omega(\Theta,\Theta)^\dagger & \Phi_{1}^{\dagger}\\ \Phi_{3} & \boxed{0}\end{vmatrix}\\
 &=
 -\begin{vmatrix}\Omega(\Theta,\Theta) 
 & \Phi_{1}^{\dagger}\\ \Phi_{3} & \boxed{0}\end{vmatrix},
\end{align}
using \eqref{skew adjoint}, and so  \eqref{SSqGr}, \eqref{SSqGr3}, and similarly \eqref{SSqGr2}, \eqref{SSqGr4}, are indeed complex conjugate.

It remains to prove that \eqref{SSqGr} and \eqref{SSqGr2} are consistent, i.e. that 
\begin{equation}
 \begin{vmatrix}\Omega(\Theta,\Theta) & \Phi_{3}^{\dagger}\\ \Phi_{1} & \boxed{0}\end{vmatrix}=
 \begin{vmatrix}\Omega(\Theta,\Theta) & \Phi_{2}^{\dagger}\\ \Phi_{3} & \boxed{0}\end{vmatrix}.
\end{equation}
Note first that this each side in this equation represents a scalar and so the right hand side can also be written as
\begin{equation}
 \begin{vmatrix}\Omega(\Theta,\Theta) & \Phi_{2}^{\dagger}\\ \Phi_{3} & \boxed{0}\end{vmatrix}^T=	
 \begin{vmatrix}\Omega(\Theta,\Theta)^T & \Phi_{3}^T\\ \Phi_{2}^* & \boxed{0}\end{vmatrix},
\end{equation}
where $^T$ denotes the matrix transpose.

Now let $\alpha$ be the $2n\times2n$ permutation matrix
\begin{equation}
\alpha=
\begin{pmatrix}
0&1&\cdots&0&0\\ 		
1&0&\cdots&0&0\\ 		
\vdots&\vdots&\ddots&\vdots&\vdots\\ 		
0&0&\cdots&0&1\\
0&0&\cdots&1&0
\end{pmatrix}. 	
\end{equation}
Pre(post)-multiplying any matrix with $2n$ (rows) columns by $\alpha$ has the effect of interchanging its $2i$th and $(2i+1)$th (rows) columns for  $i=1,\dots,n$. Hence
\begin{align}
\Phi_1\alpha=\Phi_2^*, \quad
\Phi_2\alpha=\Phi_1^*, \quad
\Phi_3\alpha=\Phi_3^*
\end{align}
and
\begin{equation}
\alpha\Omega(\Theta,\Theta)\alpha=\Omega(\Theta,\Theta)^T.
\end{equation}

Now using the quasideterminant invariance \eqref{qinv}, we get
\begin{align*}
 \begin{vmatrix}\Omega(\Theta,\Theta) & \Phi_{3}^{\dagger}\\ \Phi_{1} & \boxed{0}\end{vmatrix}&=
 \begin{vmatrix}\alpha\Omega(\Theta,\Theta)\alpha & \alpha\Phi_{3}^{\dagger}\\ \Phi_{1}\alpha & \boxed{0}\end{vmatrix}\\
 &=
 \begin{vmatrix}\Omega(\Theta,\Theta)^T& \Phi_{3}^T\\ \Phi_{2}^* & \boxed{0}\end{vmatrix}.
\end{align*}
This completes the proof.

\section{Particular solutions}\label{PSss}

In order to construct particular solutions for the Sasa-Satsuma equation \eqref{SSeqn}, we consider the quasi-Grammian solution given by \eqref{SSqGr}
\begin{eqnarray}\label{SSeqn.gs}
 u_{[n+1]}=u+2i\begin{vmatrix}\Omega(\Theta,\Theta) & \Phi_{3}^{\dagger}\\ \Phi_{1} & \boxed{0}\end{vmatrix},
\end{eqnarray}
where $\Phi_{1}$ and $\Phi_{3}$ denote the first and third rows repectively of a $3\times2n$ matrix eigenfunction $\Theta$, and the potential $\Omega(\Theta,\Theta)$ is the $2n \times 2n$ matrix with entries defined by \eqref{PotOmg} and \eqref{Omega1}--\eqref{G}. 

Let us consider the spectral problem  $L(\phi)=M(\phi)=0$ with eigenvalue $\lambda_{j}$ $(j=1,\ldots,n)$, where $\phi=\left(\phi_{3j-2}, \phi_{3j-1}, \phi_{3j}\right)^{T}$ and $L$ and $M$ are given  by \eqref{LaxL}-\eqref{LaxM} so that
\begin{eqnarray}
 \phi_{x}+J\phi\lambda_{j}+R\phi&=&0,\label{eqnL}\\
  \phi_{t}+4J\phi\lambda_{j}^{3}+4R\phi\lambda_{j}^{2}-2Q\phi\lambda_{j}+W\phi&=&0,\label{eqnM}
\end{eqnarray}
where $J$, $R$, $Q$ and $W$ are given by \eqref{JRQ}-\eqref{W}.
\vspace{3 mm}\\
$\textbf{\textit{Case (n=1)}}$
\vspace{3 mm}\\
In this case $\phi=\left(\phi_{1}, \phi_{2}, \phi_{3}\right)^{T}$ is a solution of the spectral problem $L(\phi)=M(\phi)=0$ with eigenvalue $\lambda_{1}$. Thus, from  \eqref{SSeqn.gs}, we derive the following explicit solution 
\begin{eqnarray}\label{solvacu2}
 u_{[2]}=u+2i
   \begin{vmatrix}
      F_{11} & -G_{11}^{*} & \phi_{3}^{*}\\ 
      G_{11} & F_{11} & \phi_{3}\\
      \phi_{1} & \phi_{2}^{*}& \boxed{0}
    \end{vmatrix},
\end{eqnarray}
where
\begin{eqnarray}
 F_{11}&=&\frac{1}{\lambda_{1}-\lambda_{1}^{*}}\left(\left|\phi_{1}\right|^{2}+\left|\phi_{2}\right|^{2}+\left|\phi_{3}\right|^{2}\right),\label{f}\\
 G_{11}&=&\frac{1}{2\lambda_{1}}\left(2\phi_{1}\phi_{2}+\phi_{3}^{2}\right)\label{g}.
\end{eqnarray}

\subsection{Solutions for the vacuum}

\begin{figure}
\begin{center}
 \includegraphics[width=.45\textwidth]{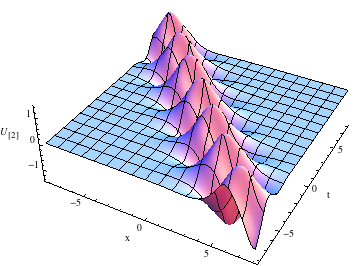}
\caption{\label{fig:1}
Breather solution $u_{[2]}$  of the SS equation \eqref{SSeqn} with the choice of parameters
$\xi=0.4, \eta=0.5$.}
\end{center}
\end{figure}

For $u=0$, the above eigenvalue problems \eqref{eqnL}-\eqref{eqnM} transform into the first-order linear system
\begin{eqnarray}
 \phi_{x}+J\phi\lambda_{j}&=&0,\\
  \phi_{t}+4J\phi\lambda_{j}^{3}&=&0,
\end{eqnarray}
which has solution $\phi=\left(\phi_{1}, \phi_{1}, \psi_{1}\right)^{T}$, where 
\begin{eqnarray}
 \phi_{j}=e^{-i\lambda_{j}\left(x+4\lambda_{j}^{2}t\right)}, \hspace{0.2cm}  \psi_{j}=e^{i\lambda_{j}\left(x+4\lambda_{j}^{2}t\right).}
\end{eqnarray}
\\
$\textbf{\textit{Case (n=1)}}$
Then, the solution \eqref{solvacu2} becomes
\begin{eqnarray}
 u_{[2]}=2i
   \begin{vmatrix}
      f & -g^{*} & \psi_{1}^{*}\\ 
      g & f & \psi_{1}\\
      \phi_{1} & \phi_{1}^{*}& \boxed{0}
    \end{vmatrix}
\end{eqnarray}
which can be written as
\begin{eqnarray}
 u_{[2]}=-2i\left(\frac{\lambda_{1}+\lambda_{1}^{*}}{\lambda_{1}-\lambda_{1}^{*}}\right)\frac{g+g^{*}}{f^{2}+|g|^{2}},
\end{eqnarray}
where
\begin{eqnarray}
 f&=&\frac{1}{\lambda_{1}-\lambda_{1}^{*}}\left(2|\phi_{1}|^{2}+|\psi_{1}|^{2}\right),\\
 g&=&\frac{1}{2\lambda_{1}}\left(2\phi_{1}^{2}+\psi_{1}^{2}\right).
\end{eqnarray}
Here the eigenfunction $\phi=\left(\phi_{1}, \phi_{1}, \psi_{1}\right)^{T}$ in which
\begin{eqnarray}
 \phi_{1}=e^{-i\lambda_{1}\left(x+4\lambda_{1}^{2}t\right)}, \hspace{0.2cm}  \psi_{1}=e^{i\lambda_{1}\left(x+4\lambda_{1}^{2}t\right).}
\end{eqnarray}
By setting $\lambda_{1}=\xi+i\eta \hspace{0.1cm}(\xi, \eta\in\mathbb{R}$ and $\xi, \eta\neq 0)$, we have
\begin{eqnarray}
 u_{[2]}=-2\frac{\xi(g+g^{*})}{\eta(f^{2}+|g|^{2})},
\end{eqnarray}
in which
\begin{eqnarray}
 f&=&\frac{1}{2i\eta}\left(2e^{2\alpha}+e^{-2\alpha}\right),\\
 g&=&\frac{1}{2(\xi+i\eta)}\left(2e^{2(\alpha-i\beta)}+e^{-2(\alpha-i\beta)}\right),
\end{eqnarray}
where $\alpha=\eta\left(x+4[3\xi^{2}-\eta^{2}]t\right)$ and  $\beta=\xi\left(x+4[\xi^{2}-3\eta^{2}]t\right)$. Then we obtain a breather solution
\begin{eqnarray}\label{SSvsol}
u_{[2]}=8\xi\eta\hspace{0.1cm}\frac{\xi\left[3\cosh(2\alpha)+\sinh(2\alpha)\right]\cos(2\beta)
              -\eta\left[\cosh(2\alpha)+3\sinh(2\alpha)\right]\sin(2\beta)}
              {\xi^{2}\left[3\cosh(2\alpha)+\sinh(2\alpha)\right]^{2}+8\eta^{2}\sin^{2}(2\beta)}.
\end{eqnarray}
The solution \eqref{SSvsol} is plotted in the Figure~\ref{fig:1}.
\subsection{Solutions for non-zero seeds}
For $u\neq0$, $u=k$ is a solution of the Sasa-Satsuma equation \eqref{SSeqn}, where we let $k$ to be a real constant. We use this as  a seed solution for application of binary Darboux transformations. Substituting $u=k$ into the linear system \eqref{eqnL}-\eqref{eqnM} and then solving for the eigenfunction $\phi=\left(\phi_{3j-2},\phi_{3j-1},\phi_{3j}\right)^{T}$, we obtain
\begin{eqnarray}
\phi_{3j-2}&=&-\frac{1}{2}c_{1}e^{-i\lambda_{j}\left(x+4\lambda_{j}^{2}t\right)}+c_{2}e^{i\sqrt{\lambda_{j}^{2}+2k^{2}}\left[x+4\left(\lambda_{j}^{2}-k^{2}\right)t\right]}+c_{3}e^{-i\sqrt{\lambda_{j}^{2}+2k^{2}}\left[x+4\left(\lambda_{j}^{2}-k^{2}\right)t\right]},\\
\phi_{3j-1}&=&\frac{1}{2}c_{1}e^{-i\lambda_{j}\left(x+4\lambda_{j}^{2}t\right)}+c_{2}e^{i\sqrt{\lambda_{j}^{2}+2k^{2}}\left[x+4\left(\lambda_{j}^{2}-k^{2}\right)t\right]}+c_{3}e^{-i\sqrt{\lambda_{j}^{2}+2k^{2}}\left[x+4\left(\lambda_{j}^{2}-k^{2}\right)t\right]},\\
\phi_{3j}&=&K_{1}e^{i\sqrt{\lambda_{j}^{2}+2k^{2}}\left[x+4\left(\lambda_{j}^{2}-k^{2}\right)t\right]}+K_{2}e^{-i\sqrt{\lambda_{j}^{2}+2k^{2}}\left[x+4\left(\lambda_{j}^{2}-k^{2}\right)t\right]},
\end{eqnarray}
where $K_{1}=i\frac{c_{2}}{k}\left(\lambda_{j}+\sqrt{\lambda_{j}^{2}+k^{2}}\right)$, $K_{2}=i\frac{c_{3}}{k}\left(\lambda_{j}-\sqrt{\lambda_{j}^{2}+k^{2}}\right)$ and $c_{1}$, $c_{2}$, $c_{3}$ are arbitrary constants.
\\\\
$\textbf{\textit{Case (n=1)}}$
The solution \eqref{solvacu2} can be written as
\begin{eqnarray}\label{u2FG}
 u_{[2]}=k-2i\frac{\left(\phi_{1}\phi_{3}^{*}+\phi_{2}^{*}\phi_{3}\right)F_{11}+\phi_{1}\phi_{3}G_{11}^{*}-\phi_{2}^{*}\phi_{3}^{*}G_{11}}{F_{11}^{2}+\left|G_{11}\right|^{2}},
\end{eqnarray}
where $F_{11}$ and $G_{11}$ are given by \eqref{f}-\eqref{g}. Substituting $\phi_{1},\phi_{2}$ and $\phi_{3}$ into \eqref{u2FG}, we obtain 
\begin{eqnarray}\label{SSu2Sol}
 u_{[2]}=k+4\mu\frac{(|\phi_{1}|^{2}-|\phi_{2}|^{2})\left(\phi_{1}\phi_{3}^{*}-\phi_{2}^{*}\phi_{3}\right)}{(|\phi_{1}|^{2}-|\phi_{2}|^{2})^{2}+2\left|\phi_{1}\phi_{3}^{*}-\phi_{2}^{*}\phi_{3}\right|^{2}},
\end{eqnarray}
where
\begin{eqnarray}
\phi_{1}&=&-\frac{1}{2}c_{1}e^{\mu\left(x-4\mu^{2}t\right)}+c_{2}e^{iD\left[x-4\left(k^{2}+\mu^{2}\right)t\right]}+c_{3}e^{-iD\left[x-4\left(k^{2}+\mu^{2}\right)t\right]},\\
\phi_{2}&=&\phi_{1}+c_{1}e^{\mu\left(x-4\mu^{2}t\right)},\\
\phi_{3}&=&-\frac{c_{2}}{k}(\mu-iD) e^{iD\left[x-4\left(k^{2}+\mu^{2}\right)t\right]}-\frac{c_{3}}{k}(\mu+iD)e^{-iD\left[x-4\left(k^{2}+\mu^{2}\right)t\right]},
\end{eqnarray}
in which $D=\sqrt{2k^{2}-\mu^{2}}$. Here, for simplicity, we have chosen $\lambda_{1} \in i\mathbb{R}$ such that $\lambda_{1}=i\mu$.

\subsubsection*{Case $D^{2}(\mu)=2k^{2}-\mu^{2}>0$}
For the case $D^{2}=2k^{2}-\mu^{2}>0$, \eqref{SSu2Sol} can be written as 
\begin{eqnarray}\label{u2p}
 u_{[2]}=k-2k\mu \frac{c^{2}K e^{2i\alpha}+{c^{*}}^{2} K^{*}e^{-2i\alpha}+2\mu|c|^{2}-2kS(ce^{i\alpha}+c^{*} e^{-i\alpha})e^{-\beta}}{\mu(c^{2}K e^{2i\alpha}+{c^{*}}^{2} K^{*}e^{-2i\alpha}) +4k^{2}|c|^{2}+2k^{2}|S|^{2} e^{-2\beta}},
\end{eqnarray}
where $\alpha=D[x-4(k^{2}+\mu^{2})t]$, $\beta=\mu(x-4\mu^{2}t)$, $K=\mu-iD$, $S=\frac{2i}{k}(|c_{2}|^{2}-|c_{3}|^{2})D$ and $c=c_{1}^{*}c_{2}+c_{1}c_{3}^{*}\neq 0$. 

\begin{figure}
\begin{center}
 \includegraphics[width=.45\textwidth]{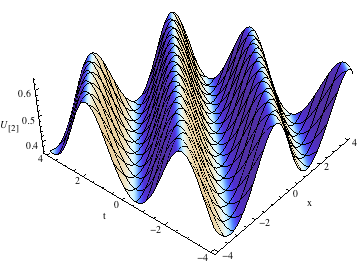}
\caption{\label{fig:2}
Periodic solution $u_{[2]}$  of the SS equation \eqref{SSeqn} with the choice of parameters
$k=0.5, \mu=0.1$.}
\end{center}
\end{figure}

\begin{figure}
\begin{center}
 \includegraphics[width=.45\textwidth]{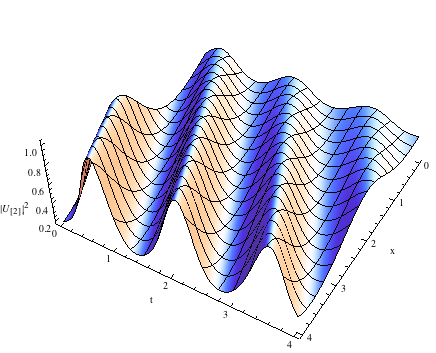}
\caption{\label{fig:3}Solution $|u_{[2]}|^{2}$  of the SS equation \eqref{SSeqn} with the choice of parameters
$k=0.7, \mu=0.3$.}
\end{center}
\end{figure}

By choosing $c_{2}=c_{3}$  so that $S=0$, we obtain a periodic solution
\begin{eqnarray}
 u_{[2]}=-k+2\frac{k(2k^{2}-\mu^{2})}{\mu\left[\mu\cos(2\alpha)+\sqrt{2k^{2}-\mu^{2}}\sin(2\alpha)\right]+2k^{2}}.
\end{eqnarray}
This solution is plotted in the Figure~\ref{fig:2}.
If we choose  $c_{1}=c_{2}-c_{3}$ then $c=|c_{2}|^{2}-|c_{3}|^{2}$ and $S=\frac{2}{k}i c D$. The solution  \eqref{u2p} can be written in the following form
\begin{eqnarray}
 u_{[2]}=k-2k\mu \frac{\mu\cos(2\alpha)+D\sin(2\alpha)+\mu-4iD[\cosh\beta -\sinh\beta]\cos\alpha}
 {\mu[\mu\cos(2\alpha)+D\sin(2\alpha)]+2k^{2}+4D^{2}[\cosh(2\beta) -\sinh(2\beta)]}.
\end{eqnarray}
This solution is plotted in the Figure~\ref{fig:3}.
\subsubsection*{Case $D^{2}(\mu)=2k^{2}-\mu^{2}<0$}
For the case $D^{2}=2k^{2}-\mu^{2}<0$, \eqref{SSu2Sol} can be written as 
\begin{eqnarray}\label{u2s}
 u_{[2]}=k-2\mu k \frac{\mu (e^{2\gamma}+ e^{-2\gamma})-E (e^{2\gamma}-e^{-2\gamma})+2\mu+4E h_{1}^{-1}h_{2}(e^{\gamma}+e^{-\gamma})e^{-\beta}}{\mu^{2} (e^{2\gamma}+ e^{-2\gamma})-\mu E (e^{2\gamma}-e^{-2\gamma})+4 k^{2}-8 E^{2}h_{1}^{-2}h_{2}^{2} e^{-2\beta}},
\end{eqnarray}
where $E=\sqrt{\mu^{2}-2k^{2}}$, $\gamma=E[x-4(k^{2}+\mu^{2})t]$, $\beta=\mu(x-4\mu^{2}t)$ and $h_{1}=c_{1}c_{2}^{*}+c_{1}^{*}c_{2}\neq 0$, $h_{2}=c_{2}c_{3}^{*}-c_{2}^{*}c_{3}$ such that $h_{1}\in\mathbb{R}$ and $h_{2}\in i \mathbb{R}$.

\begin{figure}
\begin{center}
 \includegraphics[width=.45\textwidth]{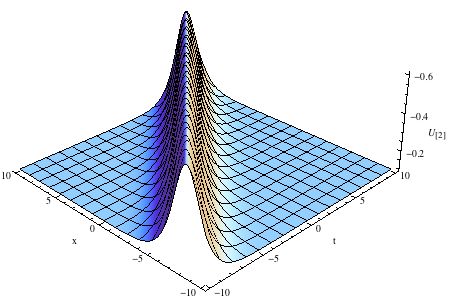}
\caption{\label{fig:4}
One-soliton solution $|u_{[2]}|^{2}$  of the SS equation \eqref{SSeqn} with the choice of parameters
$k=0.1, \mu=0.5$.}
\end{center}
\end{figure}

\begin{figure}
\begin{center}
 \includegraphics[width=.45\textwidth]{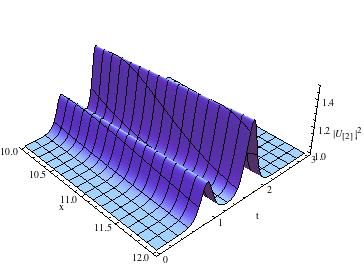}
\caption{\label{fig:5}
Two-soliton solution $|u_{[2]}|^{2}$  of the SS equation \eqref{SSeqn} with the choice of parameters
$k=1, \mu=1.5$.}
\end{center}
\end{figure}

By choosing  $c_{2}=c_{3}$  so that $h_{2}=0$, we obtain
\begin{eqnarray}
 u_{[2]}=-k-2 \frac{k(\mu^{2}-2k^{2})}{\mu^{2}\cosh(2\gamma)-\mu\sqrt{\mu^{2}-2k^{2}}\sinh(2\gamma)+2k^{2}}.
\end{eqnarray}
This is one-soliton solution. This solution is plotted in the Figure~\ref{fig:4}.
If we choose $c_{3}=-i c_{1}$ so that $h_{2}=i h_{1}$. Then the solution \eqref{u2s} can be written as
\begin{eqnarray}
 u_{[2]}=-k-2kE \frac{E (1-4 e^{-2\beta})+4 i \mu \cosh \gamma e^{-\beta}}{\mu^{2}\cosh(2\gamma)-\mu E\sinh(2\gamma)+2k^{2}+4 E^{2} e^{-2\beta}}.
\end{eqnarray}
This is two-soliton solution. This solution is plotted in the Figure~\ref{fig:5}.

\section{Conclusion}\label{ConSS}

In this paper, we have presented a standard binary Darboux transformation for the SS equation \eqref{SSeqn} and using this we have constructed a wide family of solutions in quasigrammian form. These quasigrammians  are expressed in terms of  solutions of  the linear partial differential equations given by  \eqref{LaxEqnX}-\eqref{LaxEqnT}. Moreover, periodic, one-soliton, two-solitons and breather solutions for zero and non-zero seeds have been given as particular examples for the SS equation. Examples of these solutions are plotted in the Figures~\ref{fig:1}--\ref{fig:5} for particular choices of parameters. 

One should notice that we have chosen $u=k$, where $k$ is a real constant, as a seed solution of the SS equation. This is the simplest non-zero seed. However, one might also choose the seed $u=ke^{-ia\left(x+\left[a^{2}-6k^{2}\right]t\right)}$, where $a, k\in\mathbb{R}$, to construct various rich explicit solutions including those we present here. Furthermore, in the present paper, we only consider the case $n=1$ for constructing explicit solutions. It can be obtained more explicit solutions by considering other cases such as $n=2$. Finally, it should be pointed out that the binary Darboux transformation technique is a universal instrument that allows us to construct exact solutions for other integrable systems.



\begin{thebibliography}{99}
 

 \bibitem{BNew}D. J. Benney and A. C. Newell, The propagation of nonlinear wave envelopes, \textit{J. Math. Phys}. \textbf{46}, 133--139 (1967).
 \bibitem{Ben} D. J. Benney and G. J. Roskes, Wave instabilities, \textit{Stud. Appl. Math}. \textbf{48}, 377--385 (1969).
\bibitem{CLL} H. H. Chen, Y. C. Lee and C. S. Liu, Integrability of nonlinear Hamiltonian systems
by inverse scattering method, \textit{Physica Scripta} \textbf{20}: 490--492 (1979).
\bibitem{Darboux} G. Darboux, Comptes Rendus de l'Acadmie des Sciences \textbf{94}, 1456--9 (1882).
\bibitem{Gelfand91}I. Gelfand and V. Retakh, Determinants of the matrices over noncomutative rings, Funct. Anal. App. \textbf{25}, 91--102 (1991).
\bibitem{Gelfand05}I. Gelfand, S. Gelfand, V. Retakh and R. L. Wilson, Quasideterminants, Adv. Math. \textbf{193}, 56--141 (2005).
\bibitem{GNO} C. Gilson, J. Hietarinta, J. Nimmo and Y. Ohta, Sasa-Satsuma higher-order nonlinear
Schr\"{o}dinger equation and its bilinearization and multisoliton solutions, Phys. Rev. E, \textbf{68} 016614 (2003).
\bibitem{GKundu} S. Ghosh, A. Kundu and S. Nandy, Soliton solutions, Liouville integrability and gauge equivalence of Sasa Satsuma equation, J. Math. Phys \textbf{40}, 1993 (1999). 
\bibitem{HH} B. Haider and M. Hassan, Quasi-Grammian solutions of the generalized coupled dispersionless integrable system, 
SIGMA \textbf{8 }084 (2012).
 \bibitem{Ham}M. Hamanaka, Noncommutative solitons and quasideterminants, \textit{Phys. Scr}. \textbf{89} 038006 (2014).
\bibitem{Has1}A. Hasegawa and F. Tappert, Transmission of stationary nonlinear optical pulses in
dispersive dielectric fibres I. Anomalous dispersion, \textit{Appl. Phys. Lett}. \textbf{23} 142 (1973).
\bibitem{Has2} A. Hasegawa and F. Tappert, Transmission of stationary nonlinear optical pulses in
dispersive dielectric fibres II. Normal dispersion, \textit{Appl. Phys. Lett}. \textbf{23} 171 (1973).
\bibitem{Hirota} R. Hirota, Exact envelope-soliton solutions of a nonlinear wave equation, \textit{J. Math. Phys}. \textbf{14}: 805--809 (1973).
\bibitem{KN} D. J. Kaup and A. C. Newell, An exact solution for a derivative nonlinear Schr\"{o}dinger
equation, \textit{J. Math. Phys.} \textbf{19}(4): 798--801 (1978).
\bibitem{KY} D.J. Kaup and J. Yang, The inverse scattering transform and squared eigenfunctions for a degenerate $3\times 3$ operator, Inverse Problems \textbf{25} 105010 (2009).
 \bibitem{Kiv} Y. Kivshar and G. Agrawal, \textit{Optical Solitons: From fibers to photonic crystals}, Academic
Press (2003).
 \bibitem{Kod}Y. Kodama, Optical solitons in a monomode fiber, \textit{J. Stat. Phys.} \textbf{39} 5/6 (1985).
\bibitem{KodHas}Y. Kodama and A. Hasegawa, Nonlinear pulse propagation in a monomode dielectric guide, \textit{IEEE J. Quantum Electron.} \textbf{QE-23} 510 (1987).
\bibitem{LiJon} C.X. Li and J.J.C. Nimmo, Darboux transformations for a twisted derivation and quasideterminant solutions to the super KdV equation, \textit{Proc. R. Soc. A} \textbf{466}, 2471--2493 (2009).
\bibitem{Matveev}V. B. Matveev, Darboux transformation and explicit solutions of the Kadomtcev-Petviaschvily equation, depending on functional parameters, Lett. Math. Phys. \textbf{3},  213--16 (1979).
 \bibitem{MS}V.B. Matveev and M.A. Salle, Darboux transformations and solitons, Springer Series in Nonlinear Dynamics, Springer-Verlag, Berlin (1991).
 \bibitem{NH} J.J.C. Nimmo and H. Yilmaz, On Darboux Transformations for the derivative nonlinear Schr\"{o}dinger equation, Journal of Nonlinear Mathematical Physics \textbf{21}(2) (2014) 278--293.
 \bibitem{NGO}J. J. C. Nimmo, C. R. Gilson and Y. Ohta, Applications of Darboux transformations to the self-dual Yang-Mills equations, Theor. Math. Phys.\textbf{ 122} 239--46 (2000).
 \bibitem{SS}N. Sasa and J. Satsuma, New type of soliton solutions for a higher-order nonlinear Schr\"{o}dinger equation, J. Phys. Soc. Japan \textbf{60} 409--17 (1991).
 \bibitem{Xu} T. Xu, D. Wang, M. Li and H. Liang, Soliton and breather solutions of the Sasa--Satsuma equation via the Darboux transformation, Phys. Scr.\textbf{ 89} 075207 (2014) .
  \bibitem{YK}J. Yang and D.J. Kaup, Squared eigenfunctions for the Sasa--Satsuma equation, J. Math. Phys. \textbf{50} 023504 (2009).
 \bibitem{HY}H. Yilmaz, Exact solutions of the Gerdjikov-Ivanov equation using Darboux transformations, Journal of Nonlinear Mathematical Physics \textbf{22}(1) (2015).
\bibitem{ZkSh}V. E. Zakharov and A. B. Shabat, Exact theory of two-dimensional self-focusing and
one-dimensional self-modulation of waves in nonlinear media, \textit{Zh. Eksp. Theor. Fiz}. \textbf{61} 118--134 (1971) [\textit{Sov. Phys. JETP }\textbf{34} 62--69 (1972)].
\bibitem{Zkh1} V. E. Zakharov, Stability of periodic waves of finite amplitude on the surface of a
deep fluid, \textit{Sov. Phys. J. Appl. Mech. Tech}. \textbf{4},190--194 (1968).
\bibitem{Zkh2} V. E. Zakharov, Collapse of langmuir waves, \textit{Sov. Phys. JETP} \textbf{35}, 908--914 (1972).

 
 
\end{thebibliography}
\end{document}